\documentclass{elsart}
\usepackage{amssymb}
\usepackage{amsmath}
\usepackage{pst-node}
\usepackage{graphicx}
\begin{document}

\begin{frontmatter}

%\preprint{APS/123-QED}

\title{Projective Market Model Approach to AHP Decision-Making}% Force line breaks with \\
\author[labela1]{Anna Szczypi\'nska}
\author[labela2]{, Edward W.~Piotrowski}
\address[labela1]{Institute
of Physics, University of Silesia, Uniwersytecka 4, Pl-40007 Katowice,
Poland; zanna12@wp.pl
}
 \address[labela2]{Institute
of Mathematics, University of Bia\l ystok, Lipowa 41, Pl-15424 Bia\l ystok,
Poland; ep@alpha.uwb.edu.pl
}

%\date{\today}% It is always \today, today,
             %  but any date may be explicitly specified

\begin{abstract}
In this paper, we describe market in the projective geometry language and give definition of a matrix of market rate, which is related to the matrix rate of return and the matrix of judgements in the Analytic Hierarchy Process (AHP). We use these observations to extend the AHP model to the projective geometry formalism and generalise it to an intransitive case. We give financial interpretations of such generalised model and propose its simplification. The unification of the AHP model and projective aspect of portfolio theory suggests a wide spectrum of new applications such extended model.
\end{abstract}
\begin{keyword}
AHP decision making \sep projective geometry \sep portfolio theory \sep capital processes \sep finance

\PACS 89.65.Gh \sep 02.10.Ud 
\end{keyword}
\end{frontmatter}

% PACS, the Physics and Astronomy
                             % Classification Scheme.
\section*{Introduction}
The formalism derived from algebraical rules of Venetian accountancy of projective real geometry \cite{1b} determines a natural setting for modelling the market phenomena in situations, when we can neglect the scale effects and the transaction costs. It allows us, thanks to the elegant construction of the Hilbert metric, take into account symmetries specific for modeling phenomena and analyse interesting interpretations for mathematical properties of different types of non-Euclidean geometries. The Analytic Hierarchy Process proposed by T.~L.~Saaty \cite{1a} is one of the methods of multi-criterion decision making and offers the precise quantitative method of hierarchization of criterion valuation in situations of full comparability of variants. It involves decomposing a complex decision into a hierarchy of clusters and sub-clusters, comparing properties of each possible pair of elements in each cluster as a matrix and synthesing of priorities. Therefore the AHP can be considered to be both a descriptive and a prescriptive model of decision making. It is, perhaps, the most widely used decision making method in the world today \cite{2b}. Its validity is based on thousands of actual applications in which the AHP results were accepted and used by the decision makers. We extend the AHP model to the projective geometry formalism of dual objects: price-lists and portfolios. Side effect of such financial description of the model is identification of deviation of opinion from the condition of transitivity \cite{3b, 4b} as the existence, close by articulated  hierarchy of criterions, special "transaction costs" which express aversions and preferences assessing to the concrete comparisons \cite{gifen}. We show that this feature of the model has direct connection to the matrix of capital flows used in modelling the capital changes with the help of the matrix rate \cite{5b}. It is worth to signal broad perspectives of application of such extended variant of the AHP model. This version has been successfully applied in prioritisation, resource allocation, public policy, health care, strategic planning and many more \cite{2b, gw, lgv}. The unification of the AHP model and projective aspect of portfolio theory suggests that a whole spectrum of problems can be described in such extension of the AHP model.\\
In the first section, we describe some projective markets, we define the matrix of market rate and its connection to the matrix rate of return and the matrix of judgements in the AHP model. In the second, we generalise the AHP model to intransitive case and interpret deviations from the transitivity condition in the projective market language. Finally, we discuss some method of simplification of the AHP. 

\section{Description of the projective markets}
A market, which we denote by $G$, can be a market of goods and/or obligations, criterions of judgement, information and so forth. $G$ has a natural structure of $N$-dimensional linear space over the reals \cite{1b}. Elements of this linear space are called baskets. For any basket ~$p\in G$ we have a unique decomposition into the goods which make it $$ p=\sum_{\mu=1}^N p_{\mu} \mathsf{g}_\mu $$ in some fix basis of normalised unit goods $(\mathsf{g}_1,\dots,\mathsf{g}_N)$. The element $\mathsf{g}_\mu \in G$ is the $\mu$th market good and the coefficient $p_\mu \negthinspace\negthinspace\in \negthinspace\mathbb{R}$ is called the $\mu$th coordinate of the basket and expresses its quantity. The market quotation $U$ is a linear map $U(\mathsf{g}_\nu,~\cdot~): G \rightarrow {\mathbb R}$ which assigns to every basket $p$ its current value in units of $\mathsf{g}_\nu$:\vspace{-0.5em}
$$
\label{pipi-rzut} (U p)_\nu = U(\mathsf{g}_\nu,p)  = \sum_{\mu=1}^N
U(\mathsf{g}_\nu,\mathsf{g}_\mu) p_\mu\, ,\vspace{-0.5em}
$$
where $U(\mathsf{g}_\nu,\mathsf{g}_\mu)$ is the relative price a unit of $\mu$th asset given in units of $\nu$th asset. The elements $u_{\nu\mu}:=U(\mathsf{g}_\nu,\mathsf{g}_\mu)$ for $\nu,\mu =1,\dots,N$ make a $N \times N$ matrix of market rate. 

The most general form of the matrix $u_{\nu\mu}$ was considered in paper \cite{5b}. The matrix rate of return describes the evolution of multidimensional capital and are given as a sum of two matrices: a matrix of flows, because the sum of the elements of each column is equal zero, and a matrix of growths---diagonal matrix. This description does not depend on the choice of basic goods and therefore we can consider the complex extended space of baskets and the basis of complex eigenvectors of the matrix rates in which the evolution of every capital basket can be represented as a set of noninteracting complex capital investments. In this complex basis, we do not observe any flows of the capital, but autonomous growth of individual components of the basket only. Then a description of the capital evolution is the most easily. The baskets of the complex capital have the interpretation in the real basis due to the transition matrix. In the complex extended space of baskets the decomposition of the matrix rate into the matrix of flows and the matrix of growths can always be done, in such a way that the matrix of flows is zero.

In the AHP model $u_{\nu\mu}$ is a pairwise comparison matrix of judgements. AHP involves decomposing a complex decision into a hierarchy of goal, criteria, sub-criteria and alternatives comparing properties of each possible pair of elements in each level as a matrix and synthesing of the priorities. Alternatives can be quantitative (goods) or qualitative (personal preferences). Data are collected from decision-makers in the pairwise comparison of alternatives can be based on a quantity-based judgement in kilos, metres, euro, or on a quality-based judgement---equal, strong, very strong and so on. In the second case, we have to convert these judgements into numbers, see \cite{6b}. The matrix of judgements is reciprocal $u_{\mu\nu}=(u_{\nu\mu})^{-1}$ and reflexive $u_{\nu\nu}=1$. The major problem is to find weights which order the objects and reflect the recorded judgements. Saaty suggests \cite{6b} calculating the principal right eigenvector which corresponds to the maximum eigenvalue $\lambda_{max}$ of the judgement matrix:\vspace{-0.5em}
$$\sum_{\mu=1}^{N}u_{\nu\mu}w_\mu=\lambda_{max}w_\nu\,,\;\;\;\;\; \sum_{\nu=1}^{N}w_\nu=1\,. \vspace{-0.5em}$$ The vector $w$  gives the relative weights of the alternatives. In the ideal case, when we have transitivity of the judgements $u_{\nu\rho}u_{\rho\mu}=u_{\nu\mu}$, the matrix is perfectly consistent. The entry $u_{\nu\mu}=w_\nu / w_\mu$, where $w_\nu$ is the relative weight (an unknown) of the alternative $\nu$, and the principal eigenvalue of the matrix of ratios equals~$N$. To recover the weights, we must solve the equation:\vspace{-0.5em} 
$$\;\;\; \sum_{\mu=1}^{N}u_{\nu\mu}w_\mu=N w_\nu\,.\vspace{-0.5em}$$
Real-world pairwise comparison matrices are very unlikely to be consistent and we do not have transitivity! Our thinking is intransitive because new knowledge requires that we change our minds. This is the sine qua non of progress. Moreover, the physical measurements contain errors and involve transaction costs. That is why, we generally do not have transitivity of the matrix elements. Large intransitivity unsettles our thinking, but in some situations it can be beneficial, see \cite{3b, 4b}.

In the Projective Model of Market (PMM) described in \cite{1b}, the relative prices are transitive: $u_{\nu\mu}=u_{\nu\rho}u_{\rho\mu}$. From this assumption for $\nu=\rho=\mu$ we obtain the reflexivity of the prices $u_{\nu\nu}=1$ or $u_{\nu\nu}=0$. The case $u_{\nu\nu}=0$ denotes that the $\nu$th asset is not subjected to quotation. For $\mu=\rho$ we obtain $u_{\nu\mu}u_{\mu\nu}=1$, that is antisymmetry or reciprocal. The case $u_{\nu\nu}=U(\mathsf{g}_\nu,\mathsf{g}_\nu)=1$ implies projectivity of $U$: $U((Up)_\nu\mathsf{g}_\nu)_\nu=(Up)_\nu$, because 
$$U(\mathsf{g}_\nu,(Up)_\nu\mathsf{g}_\nu)=(Up)_\nu U(\mathsf{g}_\nu,\mathsf{g}_\nu)=(Up)_\nu \,.$$
That is why, the PMM is considered. Transitivity of the relative prices denotes absence of arbitrage in the market and it is free from the scale effect.

\section{The generalised AHP model}
We consider the intransitive cases of the AHP model for which the elements of the matrix of market rate are equal $u_{\nu\mu}=\frac{w_\nu}{w_\mu}\epsilon_{\nu\mu}$, where $\epsilon_{\nu\mu}$ is the error or the noise which breaks the symmetry of the transitive models. We describe them with the help of invariables of the projective geometry. Let us assume that the prices are reflexive but not necessarily antisymmetrical, e.g. prices of currencies in the exchange offices---they contain a profit margin. We want to find the matrix $\bar{u}_{\nu \mu}:=\tfrac{q_\nu}{q_\mu} $, which is equivalent to the transitivity condition, nearest in sense of a rightly determine distance, to the matrix $u_{\nu \mu}$. A nearness measure which fulfils the assumption of PMM should be independent of scaling, that is splits of the units goods $\mathsf{g}_\nu$.  That is why, we should measure the distance in logarithms. If we choose the Frobenius norm (Euclidean), our problem will come down to minimising the functional $I(q_1,...,q_N)$, where
$$
 I(q_1,...,q_N):=||(\ln u_{\nu\mu})- (\ln \bar{u}_{\nu\mu})||^2_F=\sum_{\nu,\mu=1}^N
 (\ln u_{\nu\mu}- \ln q_\nu + \ln q_\mu)^2\,.
 $$
The scale invariant leaves one degree of freedom which is conveniently to choose so that, the prices are balanced\vspace{-0.5em} $$\sum_{\nu=0}^N \ln q_\nu = 0\;.\vspace{-0.5em}$$
Then, the functional $I(q_1,...,q_N)$ has its minimum for the prices:
$$q_\mu^\ast=\text{e}^{\sum_{\nu=1}^N \tfrac{\ln u_{\mu\nu}- \ln u_{\nu\mu}}{2N}}\,.$$
The above method to calculate the prices $q_\mu^\ast$, i.e. the transitive relative prices $\bar{u}^\ast_{\nu \mu}:=q^\ast_\nu /q^\ast_\mu$ is applied in the classic antisymmetrical version of the AHP---the Logarithmic Least Squares Method \cite{7b}, as a distance minimising method and is one of the tools for computing the priorities of the alternatives. 
The matrix of deviation from the transitivity condition $(\ln u_{\nu\mu})- (\ln \bar{u}^\ast_{\nu\mu})$ can be interpreted, in the PMM language, as the rate matrix for transaction costs (a profit margin). These profit margins destabilise the market because arbitrage is possible. When they are positive all you need is to take a short position. The generalised AHP model is applicable when the measure of such arbitrage, that is the measure of intransitivity or inconsistency, is small enough $\sqrt{I(q^\ast_1,...,q^\ast_N)}<\delta $ that this arbitrage is unprofitable. For example, it may be a result of statistics, which have errors by nature. The value of $\delta$ should be determined or adjusted by the decision-maker. This is a very important distinction. In the classic AHP model, Saaty proposes as a measure of deviation of $u_{\nu\mu}$ from the transitivity condition the consistency ratio $CR$ $$CR=\frac{\lambda_{max} -N}{(N-1)RI}\,,$$ where $N$ is order of the matrix, $\lambda_{max}$ is the maximum eigenvalue and $RI$ is an average random consistency index. Saaty suggest that $CR$ should be less than $10\%$ otherwise, we need to recalculate some of the comparisons in the matrix of judgement in order to achieve an acceptable consistency. The question is: why exactly $10\%$? This number does not follow from any mathematical model.

\section{Critical remarks. Private money in place of the AHP}
The results of deriving the relative prices $\bar{u}^\ast_{\nu \mu}$ in the generalised AHP method depend on arbitrary choice of a matrix norm, that is on a method of a distance measure from arbitrage-free situation. There is a lot of matrix norms, e.g. it is possible to generalise the above method to the Weighted Least Squares \cite{7b, 9a}. The choice of the norm is a caprice of the decision-maker and is very difficult to justify with the help of symmetries of the problem. She/he may be guided by simplicity of the solution. Nevertheless, the fact that the Logarithmic Least Squares Method can be applied with the incomplete matrices of judgement and in the case of multiple decision makers \cite{9k} is a practical advantage of that method.

In the PMM we do not have any norms but natural for market are Hilbert metrics which are defined on the projective space of portfolios or price-lists. With the help of these metrics, which have some market interpretations \cite{1b}, we can define the induced metrics. The transitive matrix that is in the nearest distance from the matrix $u_{\nu \mu}$ in such metric is unknown matrix of the market rate $\bar{u}_{\nu \mu}$. These induced metrics may have the following form: $$d_H (u_{\nu \mu},\bar{u}_{\nu \mu} ):=\max_{p\in \mathbb{R}\text{P}^N}\left( u_{\nu \mu}p, \bar{u}_{\nu \mu}p\right)_H $$ or $$d_{\text{m}}(u_{\nu \mu},\bar{u}_{\nu \mu} ):=\int_{\mathbb{R}\text{P}^N}\left( u_{\nu \mu}p, \bar{u}_{\nu \mu}p\right)_H \text{dm} (p)\,,$$ where $\mathbb{R}\text{P}^N$ is the projective space of portfolios, $\text{dm} (p)$ is a positive measure on $\mathbb{R}\text{P}^N.$ Hilbert projective metrics are particularly attractive as an indicator of closeness between two ratio scales in the AHP \cite{11} and open up new possibilities for this method.

However, collecting of the intransitive relative prices $u_{\nu\mu}$ in place of expression of the judgements straight in the form of their relative price $u_{0\mu}=U(\mathsf{g}_0 ,\mathsf{g}_\mu)$ in relation to the selected asset, to play the role of the currency $\mathsf{g}_{0}$ \cite{1b}, seem to be unnecessary throwback to the time from before 100 thousand years ago \cite{9b}, when money did not exist. Even if the currency does not take part directly in trade, suffice if each of the market entity takes itself completely private and abstract a unit currency and determines a pairwise comparisons of any type. Modern money is an abstraction too, but because it circulates between the market entities, it is an intersubjective abstraction. It will not be easier, if every decision-maker takes for $\mathsf{g}_\circledcirc$ in its mind any, not really clear for everybody else, ``coin'' and express the relative prices $(u_{\circledcirc 1},...,u_{\circledcirc N})$ of all market goods in respect of it\,? This ``coin'' is necessary to symmetrical treat of all goods which are subjected to quotation and ensures transitivity for the relative prices. Then, from the transitivity condition we obtain \vspace{-0.5em}
\begin{equation}
 \bar{u}_{\nu \mu}= u_{\nu \circledcirc} u_{\circledcirc \mu}\,.\vspace{-0.5em} \label{trans}
\end{equation}
Moreover,\vspace{-0.5em} 
\begin{equation}
1=u_{\circledcirc \circledcirc}=u_{\circledcirc \nu}u_{\nu \circledcirc}\,,\;\;\;\;\;\;\;\text{so}\;\;\;\;\;\;\;u_{\nu \circledcirc}=\frac{1}{u_{\circledcirc \nu}} \label{trans1}
\end{equation}
and we obtain antisymmetry of the relative prices. From the equations (\ref{trans}) and (\ref{trans1}), we obtain that the transitive matrix of the relative prices make the proper quotients $\bar{u}_{\nu\mu}:=\tfrac{u_{\circledcirc \mu} }{u_{\circledcirc\nu}}$. Let us observe that, the vector $(u_{\circledcirc 1},...,u_{\circledcirc N})$ gives the relative weights for the goods $1,\ldots ,N$. 
Let $(u^{(i)}_{\circledcirc_i 1},...,u^{(i)}_{\circledcirc_i N})$ be the ranking of $N$ goods coming from the $i$th decision-maker, $i=1,\ldots M$, and $a_i$ be the importance of the decision-maker in a hierarchy of decision-makers, where $\sum_{i=1}^{M}a_i =1$. Then,\vspace{-0.5em} 
$$\left(\text{\rm e}^{\sum_{i=1}^{M}a_i \ln \left( u_{\circledcirc_i 1}\right) },\ldots , \text{\rm e}^{\sum_{i=1}^{M}a_i \ln \left( u_{\circledcirc_i N}\right) }\right)\vspace{-0.5em}$$ 
is the vector of the relative weights for the $M$ decision-makers. This is the right way because only in the logarithms of the prices, additive operations of averaging transmit the transitivity property to the averaged expressions.

\section{Conclusions}
Let us observe that with the help of the vector $(u_{\circledcirc 1},...,u_{\circledcirc N})$ and transitivity, we can calculate all quotients of the matrix of the relative prices. We do not need to know all of the $\frac{N(N-1)}{2}$ parameters, but only $N$. Hence, money has the information benefit because it reduces the data $O(N^2 )\rightarrow O(N)$ (e.g. instead of controlling $5000\cong\left( \frac{100\cdot 99}{2}\right) $ parameters, we only need to operate $100$ numbers) and guarantees cost optimality, with the help of transitivity of the relative prices. The matrix of judgements is perfectly consistent and this is the very advantage of this method. The key problem in the AHP method is to remember the result of every pairwise comparison made. Practically, it is impossible for many alternatives. S.~Boz\'oki show, see \cite{15bz}, that the number of random matrices of the consistency ratio less than $10\%$ decreases dramatically as number of alternatives increases. He has generated ten million randomly matrices for every $n=3,4,\ldots, 10$ and for $n=8,9,10$, there are no matrices with $CR<10\%$.\\
The above arguments prove that, it is proper for the humans, which apply quantitative methods, to benefit from one of the oldest invention of humanity, i.e. money!

\end{document}